\documentclass[pre,twocolumn]{revtex4-2}
\usepackage{amsmath,amssymb,graphicx,dcolumn}
\usepackage[np]{numprint}
\usepackage[usenames,dvipsnames,svgnames,table]{xcolor}
\usepackage[colorlinks=true, urlcolor=Blue, linkcolor=Blue, citecolor=Blue, pdftex]{hyperref}

\npthousandsep{\,}
\npdecimalsign{.}
\npproductsign{\cdot}

\def\<{\langle}
\def\>{\rangle}

\newcolumntype{o}{D{.}{.}{1}}
\newcolumntype{t}{D{.}{.}{3}}
\newcolumntype{f}{D{.}{.}{4}}

\newcommand{\PA}{Physica}
\newcommand{\PAA}{\PA~A}

\newcommand{\JSTAT}{J.~Stat.~Mech.}

\newcommand{\PLA}{Phys.~Lett. A}

\newcommand{\PRL}{Phys.~Rev.~Lett.}

\newcommand{\JPAOLD}{J.~Phys.~A: Math.~Gen.}

\newcommand{\physrep}{Phys.~Rep.}

\newcommand{\NPB}{Nucl.~Phys.~B}
\newcommand{\JSF}{J.~Stat.~Phys.}
\newcommand{\PLB}{Phys.~Lett.~B}

\begin{document}

\title{Finite-Size scaling at fixed renormalization-group invariant}
\author{\firstname{Francesco} \surname{Parisen Toldin}}
\email{francesco.parisentoldin@physik.uni-wuerzburg.de}
\affiliation{\mbox{Institut f\"ur Theoretische Physik und Astrophysik, Universit\"at W\"urzburg, Am Hubland, D-97074 W\"urzburg, Germany}}
\begin{abstract}
  Finite-size scaling at fixed renormalization-group invariant is a powerful and flexible technique to analyze Monte Carlo data at a critical point.
  It consists in fixing a given renormalization-group invariant quantity to a given value, thereby trading its statistical fluctuations with those of a parameter driving the transition.
  One remarkable feature is the observed significant improvement of statistical accuracy of various quantities, as compared to a standard analysis.
  We review the method, discussing in detail its implementation, the error analysis, and a previously introduced covariance-based optimization.
  Comprehensive benchmarks on the Ising model in two and three dimensions show large gains in the statistical accuracy, which are due to cross-correlations between observables.
  As an application, we compute an accurate estimate of the inverse critical temperature of the improved O(2) $\phi^4$ model on a three-dimensional cubic lattice.
\end{abstract}

\maketitle

\section{Introduction}
\label{sec:intro}
One of the most powerful techniques in numerical studies of systems with a large number of degrees of freedom is the (dynamic) Monte Carlo (MC) method \cite{Sokal_lecture}.
It essentially consists in the realization of a Markov chain, which at equilibrium converges to a target distribution, typically the Gibbs measure.
Due to its flexibility, MC methods are commonly used in numerical studies of critical phenomena \cite{PV-02}.
Since truly singular behavior is found in the thermodynamic limit only, one needs extrapolation techniques in order to extract the critical behavior from numerical simulations which necessarily deal with a finite number of degrees of freedom.

For this purpose, one of the simplest approaches consists in extrapolating the infinite-volume limit from MC data at a finite size $L$.
In the presence of a finite exponential correlation length $\xi$, and in the grand-canonical ensemble, observables generically exhibit an approach to their thermodynamic limit which is exponential in the ratio $L/\xi$,
i.e., their finite-size correction is proportional to $\exp(-L/\xi)$, for $\xi\ll L$ \cite{Luescher-86,Neuberger-89,Muenster-85}.
Nevertheless, finite-size corrections polynomial in $1/L$ are found for some specific observables, such as the second-moment correlation length \cite{CP-98,PTHAH-14}, or in the absence of translational invariance; for instance, with open boundary conditions (BCs), subleading terms in the free energy give rise to corrections proportional to $1/L$.
Interestingly, the statement of exponential approach to the infinite-volume limit does not hold in the canonical ensemble, where the constraint on the particle number induces finite-size corrections proportional to the inverse volume \cite{WAPT-17}.

In the vicinity of a critical point, due to the divergence of the correlation length $\xi$, infinite-volume methods have a limited application.
In this context, finite-size scaling (FSS) techniques play an important role \cite{Privman-90,PV-02}.
FSS is formulated by applying renormalization-group (RG) theory
to a system of finite size, and close to a critical point.
This leads to the prediction of a scaling form for the singular part of the free energy, and for derived quantities, such that they acquire a dependence on the ratio $\xi/L$.
This scaling behavior can be then exploited to formulate specific FSS methods; we refer to Refs.~\cite{Privman-90,PV-02} for a discussion of the various techniques.
The great merit of FSS lies in the fact that,
unlike infinite-volume methods mentioned above, it works in a parameter region where $\xi$ is of the order of the linear size of the system $L$.
For this reason, FSS is very often the method of choice for numerical studies of critical phenomena.
In FSS a central role is played by the so-called RG-invariant observables.
These are quantities which are scale-invariant at a critical point, and are commonly used to locate its position \cite{Privman-90,PV-02}.

In the method of FSS at fixed RG-invariant, also known as FSS at fixed phenomenological coupling \cite{Hasenbusch-99}, one fixes the value of a chosen RG-invariant quantity $R$, thereby trading the statistical fluctuations of $R$ with those of a parameter driving the transition.
This method, introduced in Ref.~\cite{Hasenbusch-99}, does not require additional computational time and has been used in several high-precision MC studies \cite{HPV-05,CHPV-06,HPTPV-07,HPTPV-07b,HPTPV-08b,Wolff-09c,Hasenbusch-10,Hasenbusch-19,Hasenbusch-20,PT-20}.
It shares some similarities with the method of phenomenological renormalization, where two system sizes are enforced to share a common value of $R=\xi/L$ \cite{Nightingale-76,Nightingale-77}.
One interesting feature of FSS at fixed RG-invariant is that, compared to a standard analysis, in some cases a significant reduction of error bars has been reported \cite{HPV-05,HPTPV-07,Wolff-09c,PT-11}.
Considering that, for an equilibrated Markov chain, the accuracy of sampled observables is generically proportional to the inverse square root of the number of MC samples, even a (relatively) small improvement in the error bars may represent a significant gain in terms of computational time.
This aspect is particularly important in the context of critical phenomena because of the occurrence of critical slowing down \cite{Sokal_lecture}.

In Ref.~\cite{PT-11} we have introduced an optimization of FSS at fixed RG-invariant which aims at minimizing the error bars of MC observables by determining the optimal RG-invariant to be fixed, as a linear combination of available RG-invariants.
This is accomplished by an analysis of the statistical covariance between observables, and, as benchmarked in Ref.~\cite{PT-11}, leads to large gains in CPU time.

In this work we give a comprehensive overview of the method of FSS at fixed RG-invariant, benchmarking its performance in improving the statistical error bars.
As a concrete application, we determine the location of the critical point in the improved O(2) $\phi^4$ lattice model, whose critical behavior belongs to the three-dimensional $XY$ universality class.

This paper is organized as follows.
In Sec.~\ref{sec:fss} we discuss the formulation of the method, paying particular attention to its practical implementation, and to the error analysis;
 we also describe in more detail the covariance optimization outlined in Ref.~\cite{PT-11}.
In Sec.~\ref{sec:benchmarks} we report extensive benchmarks of the method on the Ising model in two and three dimensions.
In Sec.~\ref{sec:xy} we apply FSS at fixed RG-invariant to determine the critical inverse temperature of the $\phi^4$ lattice model with two components in three dimensions.
A summary of the present work is given in Sec.~\ref{sec:summary}.
In the Appendix we give the proof of two formulas for the coefficients of an optimized linear combination of RG-invariants.

\section{FSS at fixed RG-invariant}
\label{sec:fss}

\subsection{Formulation of the method}
\label{sec:fss:method}
A common approach in FSS methods \cite{Privman-90} is the notion of a pseudocritical coupling.
Generically, this is the (apparent) value of the coupling constant at the onset of criticality, as determined at a {\it finite} size $L$.
Since no true critical point can appear in a finite size, the actual definition is not unique.
For example, a pseudocritical coupling can be defined as the maximum of the susceptibility of the order parameter for a finite system.
Another popular definition is provided by the crossing method: one defines a pseudocritical coupling by the intersection of the curves of an RG-invariant observable at size $L$ and size $\alpha L$, with fixed $\alpha > 1$ (see, e.g., Ref.~\cite{PTHAH-14}).
By construction, pseudocritical couplings converge to the true critical-point constant for $L\rightarrow \infty$.

For concreteness, here and in the following we employ the language of classical critical phenomena at finite temperature, where the tuning parameter is the inverse temperature $\beta$;
the method discussed in the following can be readily applied also to quantum critical phenomena, where FSS holds and the tuning parameter is a coupling constant present in the Hamiltonian \cite{CPV-14}.
We define a pseudocritical inverse temperature $\beta_f(L)$ by fixing a given RG-invariant quantity $R$ to a fixed value $R_f$:
\begin{equation}
  R(\beta_f(L), L) = R_f.
  \label{Rfixed}
\end{equation}
According to RG theory and FSS, in the vicinity of a critical point located at $\beta=\beta_c$,
\begin{equation}
  R(\beta, L) = f(t L^{1/\nu}) + L^{-\omega}g(t L^{1/\nu}),
  \label{fss_R}
\end{equation}
where $1/\nu$ is the RG dimension of the relevant, non-symmetry-breaking, scaling field $t = a_0(\beta-\beta_c)/\beta_c$, and $\omega$ is the leading correction-to-scaling exponent.
The scaling function $f(z)$ is universal, once the normalization $a_0$ on the argument $z$, the shape, and BCs are fixed \cite{KOK-99,PTD-10,HGS-11,MFG-14}.
We indicate with $R^*\equiv f(0)$ the critical-point value of $R$.
Inserting Eq.~(\ref{fss_R}) in Eq.~(\ref{Rfixed}) and solving it for $z$ one finds for $L\rightarrow\infty$
\begin{equation}
  \begin{split}
    z &= a_0\left(\frac{\beta_f(L)-\beta_c}{\beta_c}\right)L^{1/\nu}\\
    &\simeq f^{-1}(R_f) - L^{-\omega}\frac{g(f^{-1}(R_f))}{f'(f^{-1}(R_f))}.
  \end{split}
  \label{Rfixed_solution}
\end{equation}
From Eq.~(\ref{Rfixed_solution}) it follows that the pseudocritical inverse temperature $\beta_f(L)$ converges to $\beta_c$ as
\begin{align}
  \label{betaf_vs_betac_gen}
  \beta_f(L)-\beta_c &\propto L^{-1/\nu}, \qquad &\text{generic $R_f$},\\
  \label{betaf_vs_betac_crit}
  \beta_f(L)-\beta_c &\propto L^{-1/\nu-\omega}, \qquad &R_f=R^*,
\end{align}
i.e., the convergence of $\beta_f(L)$ is improved when we fix $R$ to its critical-point value $R^*$.
Once $\beta_f(L)$ is determined as a solution of Eq.~(\ref{Rfixed}), the various observables are analyzed at $\beta=\beta_f(L)$, so that they depend on $L$ only.
In particular, let us consider a divergent observable $O$ that in the FSS limit satisfies a scaling ansatz of the form
\begin{equation}
  O(\beta, L) = L^e\left[f_O(t L^{1/\nu}) + L^{-\omega}g_O(t L^{1/\nu})\right],
  \label{fss_obs}
\end{equation}
where $e$ is a critical exponent
\footnote{In Eq.~(\ref{fss_obs}) we have for simplicity assumed that the leading correction-to-scaling in $O$ decays with the same exponent $\omega$ appearing in the Eq.~(\ref{fss_R}).
This is indeed the case when the dominant corrections are given by the leading irrelevant scaling field, which generically enters in the FSS limit of all observables.
Nevertheless, for some specific observables the leading scaling correction has another origin, for example it could be due to the nonsingular part of the free energy.
This is the case of the susceptibility in the Ising model on the square lattice\cite{PV-02}, or of the zero-momentum correlations of the order parameter in the ground state of the Hubbard model on the honeycomb lattice, at the onset of the quantum phase transition between a semimetallic and an antiferromagnetic phase \cite{PTHAH-14}.
In such cases, the leading correction-to-scaling exponent $\omega$ entering in Eq.~(\ref{fss_obs_fixedR}) is the smallest between those appearing in Eq.~(\ref{fss_R}) and Eq.~(\ref{fss_obs}).}.
Using Eq.~(\ref{Rfixed_solution}) in Eq.~(\ref{fss_obs}), we obtain the behavior of $O$ at fixed $R=R_f$,
\begin{equation}
  O_f(L)\equiv O(\beta_f(L), L) = A L^e \left(1+BL^{-\omega}\right),
  \label{fss_obs_fixedR}
\end{equation}
where $A$ and $B$ are observable-dependent constants.
We notice that Eq.~(\ref{fss_obs_fixedR}) holds irrespectively of the choice of $R_f$.
In other words, unlike Eq.~(\ref{betaf_vs_betac_crit}), the choice of $R_f=R^*$ does not suppress corrections in $O$ at fixed $R=R_f$.
Equation (\ref{fss_obs_fixedR}) is the starting point to determine the critical exponent $e$ by fitting the size dependence of $O_f$.
In Sec.~\ref{sec:benchmarks} we consider as observables the susceptibility $\chi$ and the derivatives of RG-invariant quantities $dR/d\beta$, as estimators for the critical exponents $\eta$ and $\nu$.

\subsection{Implementation of the method}
\label{sec:fss:implementation}
Typically, MC simulations are performed at a fixed value of $\beta=\beta_{\rm run}$.
The implementation of the method requires to extrapolate the observables at a different value of $\beta$, so to be able to solve Eq.~(\ref{Rfixed}) and to compute Eq.~(\ref{fss_obs_fixedR}).
For this purpose, one can sample the derivatives of observables with respect to $\beta$, and compute a Taylor expansion
\begin{equation}
  O(\beta) \simeq O(\beta_{\rm run}) + \frac{\partial O}{\partial\beta}(\beta-\beta_{\rm run}) + \frac{1}{2}\frac{\partial^2 O}{\partial\beta^2}(\beta-\beta_{\rm run})^2,
  \label{taylor}
\end{equation}
where $O(\beta_{\rm run})$ indicates an observable $O$ as sampled via MC simulation at $\beta=\beta_{\rm run}$.
Generally, a second-order expansion as in Eq.~(\ref{taylor}) should be sufficient to reliably extrapolate $O$ to values of $\beta$ close to the simulation parameter $\beta_{\rm run}$.
Nevertheless,
it is advisable to repeat the calculations truncating the expansion to the first and second term in Eq.~(\ref{taylor}), so as to to check the stability of the extrapolation.
Alternatively, one can use reweighting techniques to compute $O(\beta)$.

An important aspect of the method is the correct determination of the uncertainty of observables $O_f$ at fixed RG-invariant $R$.
For this purpose, one can use the jackknife method \cite{Young_notes}.
This consists in dividing the MC measures into $N_{\rm bin}$ bins of equal size, after having discarded the initial measures to ensure the equilibration of the Markov chain.
The size of each bin needs to be larger than the autocorrelation time of the various observables, such that the bins can be regarded as statistically independent from each other.
Then, one defines a jackknife bin $j$ as the set of MC measures of all bins, excluded the bin $j$.
For each jackknife bin $j$, one can compute the various observables, in particular the chosen RG-invariant quantity $R$.
This results in $N_{\rm bin}$ jackknife estimates, which we denote with $R^{(j)}$.
Employing the Taylor expansion of Eq.~(\ref{taylor}) (or a reweighting method) in Eq.~(\ref{Rfixed}), one determines the $j-$th jackknife estimate of $\beta_f^{(j)}$.
For a generic observable $O$, this is used in Eq.~(\ref{taylor}) to compute $O_f$, the value of $O$ at fixed RG-invariant.
Repeating this procedure for all $N_{\rm bin}$ jackknife bins results in  $N_{\rm bin}$ estimates of $O_f^{(j)}$.
Furthermore, we also consider the value of $O_f$, as obtained using the entire set of (equilibrated) MC data, and denote it with $O_f^{({\rm all})}$.
The jackknife estimators for the central value of $O_f$ and its variance are \cite{Young_notes}
\begin{align}
  \label{jackknife_mean}
  O_f^{\rm est} &= N_{\rm bin}O_f^{({\rm all})} - \left(N_{\rm bin}-1\right)\overline{O_f^{(j)}},\\
  \label{jackknife_sigma}
  \left(\sigma_{O_f}^2\right)^{\rm est} &= \frac{N_{\rm bin}-1}{N_{\rm bin}}\sum_{j=1}^{N_{\rm bin}}\left(O_f^{(j)}-\overline{O_f^{(j)}}\right)^2,
\end{align}
where $\overline{O_f^{(j)}}$ is the average over the jackknife bins
\begin{equation}
  \overline{O_f^{(j)}} \equiv  \frac{1}{N_{\rm bin}}\sum_{j=1}^{N_{\rm bin}}O_f^{(j)}.
  \label{average_bins}
\end{equation}
Similarly, the jackknife method offers an estimator for the covariance between two observables $A$ and $B$,
\begin{multline}
    {\rm COV}(A, B)^{\rm est} \\
    = \frac{N_{\rm bin}-1}{N_{\rm bin}}\sum_{j=1}^{N_{\rm bin}}\left(A^{(j)}-\overline{A^{(j)}}\right)\left(B^{(j)}-\overline{B^{(j)}}\right),
  \label{jackknife_covariance}
\end{multline}
where, analogous to the discussion above, $A^{(j)}$, $B^{(j)}$ are the estimates of $A$, $B$ on the jackknife bin $j$, and $\overline{A^{(j)}}$, $\overline{B^{(j)}}$ their average [see Eq.~(\ref{average_bins})].

Alternative methods for estimating error bars are the bootstrap \cite{Young_notes} and the $\Gamma-$method \cite{Wolff-04,*Wolff-04_erratum}.

We end this section with a few technical remarks.
As observed above,
the determination of the pseudocritical inverse temperature $\beta_f$ and of the observables at fixed RG-invariant $O_f$ requires an extrapolation of MC data sampled at a given value of $\beta$.
Irrespective of the method employed [Taylor expansion as in Eq.~(\ref{taylor}) or a reweighting technique], such an extrapolation can  only be reliably performed for values of $\beta$ close enough to $\beta_{\rm run}$.
If in performing this analysis the value of $\beta_f$ is significantly distant from $\beta_{\rm run}$, one may need to redo the simulation at value of $\beta$ closer to the determined $\beta_f$.
As a practical recipe,
one can start from a small lattice size $L_0$, where repeating a simulation does not incur in a significant overhead, determining $\beta_f(L_0)$.
This value can be then employed as $\beta_{\rm run}$ for the simulation of the next lattice size $L_1 > L_0$.
Like before, if the value of $\beta_f(L_1)$ found differs significantly from $\beta_{\rm run}$, the simulation may need to be repeated at a value of $\beta_{\rm run}$ closer to the estimated $\beta_f(L_1)$.
Following this procedure, one can systematically increase the lattice size, determining $\beta_f(L)$ and $O_f(L)$.
Equations (\ref{betaf_vs_betac_gen}) and (\ref{betaf_vs_betac_crit}) ensure a quick convergence of $\beta_f(L)$, so that the overhead of repeated simulations is limited to the smaller lattice sizes only.

As discussed in Sec.~\ref{sec:fss:method}, the fixed value $R_f$ can, in principle, be arbitrary.
In practice, however, it is preferable to choose $R_f$ as close as possible to the critical-point value $R^*$.
There are at least three reasons for this.
First of all, as shown in Eq.~(\ref{betaf_vs_betac_crit}), when $R=R^*$, there is an improved convergence of $\beta_f(L)$ to $\beta_c$.
Second, even if we choose a value $R_f\ne R^*$, the prefactor in front of the right-hand side of Eq.~(\ref{betaf_vs_betac_gen}) is proportional to $R_f - R^*$, such that a value of $R_f$ significantly different from the critical-value $R^*$ gives rise to a (numerically) slower approach of $\beta_f(L)$ to $\beta_c$ (although the asymptotic power-law exponent $1/\nu$ remains independent of $R_f$ when $R_f\ne R^*$).
Third, the analysis of Sec.~\ref{sec:fss:method} considers only the leading correction-to-scaling term.
For a value of $R_f$ sufficiently far from $R^*$, further subleading terms in Eq.~(\ref{fss_R}) may be relevant and slow the approach to the FSS limit.

In a given model,
when an estimate of $R^*$ is not available from previous results, a common approach is to use standard FSS technique to determine the value of $R^*$, for example by using crossing methods, or by fitting a Taylor expansion of the right-hand size of Eq.~(\ref{fss_R}) \cite{Hasenbusch-10,Hasenbusch-19,Hasenbusch-20,PT-20}.
For this purpose one does not generally need large lattice sizes.
Once an estimate of $R^*$ is determined, one can proceed to analyze the MC data collected so far at $R=R^*$, possibly supplementing them with additional data obtained from simulations at larger lattice sizes.
In this way, one optimizes the use of CPU time, reserving most of it for simulations at a single value of $\beta_{\rm run}$ and large lattice sizes.

\subsection{Error analysis}
\label{sec:fss:error}
MC simulations are a realization of a Markov chain, whose equilibrium distribution is the target Boltzmann-Gibbs measure \cite{Sokal_lecture}.
The outcome of the simulations, and thus
the various statistical estimators discussed in the previous section, are actually stochastic variables, whose expectation value is equal to the quantity of interest.
We indicate with $E[X]$ the expectation value of a stochastic variable $X$, and its fluctuations around the mean value with $\delta X \equiv X - E[X]$.
As discussed in Sec.~\ref{sec:fss:implementation}, Eq.~(\ref{Rfixed}) can be reliably solved only if the simulation parameter $\beta_{\rm run}$ is sufficiently close to $\beta_f(L)$.
In this case, to the leading order in the Taylor expansion of an RG-invariant quantity, we have
\begin{equation}
  \beta_f(L) \simeq \beta_{\rm run} + \frac{R_f - R(\beta_{\rm run}, L)}{\frac{\partial R}{\partial \beta}\left(\beta_{\rm run}, L\right)}.
  \label{solution_betaf}
\end{equation}
By expanding Eq.~(\ref{solution_betaf}) around the expectation value, and keeping the lowest order in the fluctuations, we obtain
the fluctuations of $\beta_f$,
\begin{equation}
    \delta \beta_f \simeq -\frac{\delta R}{E\left[\frac{\partial R}{\partial \beta}\right]}
    + \frac{R_f - E\left[R\right]}{E\left[\frac{\partial R}{\partial \beta}\right]^2}\delta \frac{\partial R}{\partial \beta},
  \label{fluctuations_betaf}
\end{equation}
where to keep a light notation, observables on the right-hand side are meant to be computed at the parameter of the simulation $\beta_{\rm run}$, and at a given lattice size $L$.
The second term on the right-hand side of Eq.~(\ref{fluctuations_betaf}) represents a correction and, according to the discussion above, we can neglect it as long as $R\simeq R_f$.
The variance of $\beta_f(L)$ is then
\begin{equation}
  \sigma_{\beta_f}^2 = \frac{\sigma_R^2}{E\left[\frac{\partial R}{\partial \beta}\right]^2}.
  \label{var_betaf}
\end{equation}
Given an observable $O$, its value at fixed $R=R_f$ can be computed by inserting Eq.~(\ref{solution_betaf}) in the Taylor expansion of Eq.~(\ref{taylor}). To the lowest order in $\beta_{\rm run}-\beta_f(L)$, we have
\begin{equation}
  O_f(L) \simeq O(\beta_{\rm run}, L) + \frac{\partial O}{\partial\beta}\frac{R_f - R(\beta_{\rm run}, L)}{\frac{\partial R}{\partial \beta}\left(\beta_{\rm run}, L\right)}.
  \label{solution_O_fixedR}
\end{equation}
Expanding around the expectation value, we have, to the lowest order in the fluctuations,
\begin{equation}
  \begin{split}
    \delta O_f \simeq \delta O &- E\left[\frac{\partial O}{\partial\beta}\right] \frac{\delta R}{E\left[\frac{\partial R}{\partial \beta}\right]}
    +\delta \frac{\partial O}{\partial\beta}\frac{R_f - E\left[R\right]}{E\left[\frac{\partial R}{\partial \beta}\right]} \\
    & - E\left[\frac{\partial O}{\partial\beta}\right]\frac{R_f - E\left[R\right]}{E\left[\frac{\partial R}{\partial \beta}\right]^2}\delta\frac{\partial R}{\partial \beta}.
  \end{split}
  \label{fluctuations_Of}
\end{equation}
Similar to Eq.~(\ref{fluctuations_betaf}), the second and the third term on the right-hand side of Eq.~(\ref{fluctuations_Of}) can be neglected in first approximation, leading to the variance of $O_f$:
\begin{equation}
  \sigma_{O_f}^2 = \sigma_{O}^2 + \frac{E\left[\frac{\partial O}{\partial\beta}\right]^2}{E\left[\frac{\partial R}{\partial \beta}\right]^2}\sigma_R^2 - 2 \frac{E\left[\frac{\partial O}{\partial\beta}\right]}{E\left[\frac{\partial R}{\partial \beta}\right]}{\rm COV}(O, R).
  \label{var_Of}
\end{equation}
The variance of an observable $O$ at fixed $R=R_f$ is thus a sum of the variance of the observable $O$, i.e., the one arising from a standard analysis at fixed Hamiltonian parameters, and two additional terms. The first term is due to the fluctuations of the RG-invariant quantity $R$ and it is always positive.
The second term, proportional to the covariance between $O$ and $R$, is the term that is, potentially, responsible for the reduction of error bars.
It is worth noticing that if there is little covariance between $O$ and $R$, the final uncertainty of $O$ at fixed $R=R_f$ is actually larger than the standard one.
In fact, the large covariance found between observables and RG-invariant quantity is the origin of the drastic reduction of error bars reported in some cases \cite{HPTPV-07,PT-11}.

\subsection{Covariance optimization}
\label{sec:fss:covariance}
In Ref.~\cite{PT-11} we have proposed a method to optimize the FSS at fixed RG-invariant, by minimizing the error bar of the resulting observable $O_f$.
For this purpose, one considers a set of RG-invariant observables $\{R_i\}$ and a linear combination of them,
\begin{equation}
  R(\{\lambda_i\}) \equiv \sum_i \lambda_i R_i.
  \label{R_lambda}
\end{equation}
We search in the space of linear combinations of $\{R_i\}$, the optimal coefficients $\lambda_i$ that minimize the variance of an observable $O$ at fixed $R(\{\lambda_i\})$.
The solution to this problem is, up to an inessential normalization of $\{\lambda_i\}$ \cite{PT-11},
\begin{equation}
\lambda_i = -\frac{{\bf R'^T M^{-1}N}-E[\partial O/\partial\beta]}{\bf R'^TM^{-1}R'}\left({\bf M^{-1}R'}\right)_i+\left({\bf M^{-1}N}\right)_i,
\label{optimal_lambda_O}
\end{equation}
where the matrix ${\bf M}$ and the vectors ${\bf N}$ and ${\bf R'}$ are defined as
\begin{equation}
  \begin{split}
    {\bf M}_{ij} \equiv {\rm COV}(R_i, R_j),\\
    {\bf N}_i \equiv {\rm COV}(O, R_i),\\
    {\bf R'}_i \equiv E\left[\frac{\partial R_i}{\partial\beta}\right].
  \end{split}
  \label{def_MNRp}
\end{equation}
Analogously, one can consider the problem of finding the optimal linear combinations of $\{R_i\}$ that minimizes the variance of $\beta_f$.
In this case, the solution is
\begin{equation}
  \lambda_i = \frac{\left({\bf M^{-1}R'}\right)_i}{\bf R'^TM^{-1}R'}.
  \label{optimal_lambda_betaf}
\end{equation}
with ${\bf M}$ and ${\bf R'}$ as in Eq.~(\ref{def_MNRp}).
In the Appendix we prove Eq.~(\ref{optimal_lambda_O}), whose proof was only sketched in Ref.~\cite{PT-11}, as well as Eq.~(\ref{optimal_lambda_betaf}).
The covariances in the definitions of Eq.~(\ref{def_MNRp}) can be estimated within the jackknife method using Eq.~(\ref{jackknife_covariance}).
Covariances and estimated error bars are in general related to the transition matrix defining the Markov chain \cite{Sokal_lecture}.
To illustrate this aspect, in Sec.~\ref{sec:benchmarks} we test the method using different sampling algorithms.

In implementing the covariance-optimized FSS at fixed RG-invariant, it is important to note that the chosen RG-invariant observable $R$ must be the same for all lattice sizes $L$.
On the other hand, the optimal choice of $\{\lambda_i\}$ obtained from Eq.~(\ref{optimal_lambda_O}) does in general depend on $L$.
Thus, a possible strategy is to choose the values of $\{\lambda_i\}$ that minimize the error bar on $O$ for the largest available lattice size, where typically the most part of the CPU time is consumed, and then use the resulting RG-invariant in the FSS analysis of all lattice sizes.
On the other hand, it is possible (and useful) to optimize $\{\lambda_i\}$ differently for each observable $O$ considered, as long as one does not combine different observables in the scaling analysis.
This is the strategy that we used in Ref.~\cite{PT-11} to test the covariance optimization.

\section{Benchmarks}
\label{sec:benchmarks}
\subsection{Model and observables}
\label{sec:benchmarks:model}
To benchmark the method, we have simulated the standard Ising model, whose Hamiltonian on a $d-$dimensional lattice is
\begin{equation}
\label{ising}
{\cal H} = -J\sum_{\<ij\>}\sigma_i\sigma_j,\qquad \sigma_i=\pm 1,
\end{equation}
where the sum is over the nearest-neighbors sites. Without loss of generality, we set $J=1$.
We apply periodic BCs on a finite lattice with linear extension $L$.
To test the influence of the sampling algorithm on the performance of the method, we have simulated the model using either the standard Metropolis, or the Wolff single-cluster updates \cite{Wolff-89}.
Both are ergodic algorithms for the Ising model \cite{Sokal_lecture}.

Here we consider four RG-invariant observables: $\xi/L$, $U_4$, $U_6$, and $Z_a/Z_p$, defined as follows.
$\xi$ is the second-moment correlation length at a finite size, defined as
\begin{equation}
  \xi = \frac{1}{2\sin \left(\hat{p}/2\right)}\sqrt{\frac{C(0)}{C(\hat{p})}-1},
  \label{xi}
\end{equation}
where $C(p)$ is the Fourier transform of the two-point function of the order parameter $\sigma_i$, and $\hat{p} = 2\pi/L$ is the minimum nonzero momentum on a finite lattice with size $L$;
see Appendix A of Ref.~\cite{PTHAH-14} for a discussion on the definition of $\xi$ in a finite lattice.
The generalized Binder ratios $U_4$ and $U_6$ are defined by
\begin{equation}
  U_{2j} \equiv \frac{\<M^{2j}\>}{\<M^2\>^j},
  \label{binder}
\end{equation}
where $M\equiv \sum_i\sigma_i$ is the total magnetization of the model, and the brackets $\<\quad\>$ indicate the thermal average.
Finally, $Z_a/Z_p$ is the ratio of the partition function of the model with antiperiodic BCs on one direction and periodic BCs on the remaining ones, over the partition function with full periodic BCs.
This quantity can be efficiently sampled with the boundary-flip algorithm of Ref.~\cite{Hasenbusch-93,Hasenbusch-01}.

As for the observables $O$ which we analyze at fixed RG-invariant, we consider here the susceptibility $\chi$
\begin{equation}
  \chi \equiv \frac{1}{L^d} \sum_{i j} \sigma_i \sigma_j,
  \label{chi}
\end{equation}
which in the FSS limit is $\propto L^{2-\eta}$, thus serving as an estimator of the critical exponent $\eta$.
We also compute the first derivative of the RG-invariant quantities defined above with respect to $\beta$: $d(\xi/L)/d\beta$, $dU_4/d\beta$, $dU_6/d\beta$, and $d(Z_a/Z_p)/d\beta$. These are $\propto L^{1/\nu}$ in the FSS limit and are commonly used as estimators of the critical exponent $\nu$ \cite{Privman-90}.

\subsection{Results}
\label{sec:benchmarks:results}
We have simulated the Ising model in $d=2$, at the critical point $\beta_{\rm run}=\numprint{0.4406867935}$ \cite{baxter-book}, and for lattice sizes $L=8$-$128$.
The analysis has been carried out using the fixed-point values of the RG-invariants, which thanks to the available exact solutions, are known to a high numerical accuracy.
We use $(\xi/L)^*=0.9050488292(4)$, $U_4^*=1.167923(5)$, $U_6^*=1.455649(7)$, \cite{SS-00}, and $(Z_a/Z_p)^*\simeq 0.37288488$ \cite{CH-96}.
For each observable, we compute its value fixing one of the RG-invariants.
Since in the MC method the uncertainty of an observable is $\propto 1/\sqrt{N_{\rm steps}}$, with $N_{\rm steps}$ the number of MC steps in the sampling, the effective gain in CPU time can be defined as
\begin{equation}
  \text{CPU gain} = \left(\frac{\sigma_\text{std}}{\sigma_\text{fixed-RG}}\right)^2,
  \label{cpugain}
\end{equation}
where $\sigma_\text{std}$, $\sigma_\text{fixed-RG}$ are the error bar estimates for a standard analysis, and for the analysis at fixed RG-invariant.
For comparison we also show the CPU gains obtained using the covariance-optimized linear combination of RG-invariants; for observables $\chi$, $d(\xi / L)/d\beta$, $dU_4/d\beta$, $dU_6/d\beta$ these are taken from Ref.~\cite{PT-11}.
As discussed in Sec.~\ref{sec:fss:covariance}, for each observable, the coefficients of the linear combination are fixed by the minimization of the resulting error bar for the largest lattice size.

\begin{table}\renewcommand*{\arraystretch}{1.4}
  \caption{Benchmark results for the 2D Ising model, sampled with a Metropolis dynamics.
    We report the effective gain in CPU time [Eq.~(\ref{cpugain})] in the analysis of various observables at fixed RG-invariant $R$, as compared to the standard analysis at fixed MC parameters.
    For each lattice size $L$, we compare the gains obtained by choosing $R=\xi/L$, $U_4$, $U_6$, $Z_a/Z_p$, and by considering the covariance-optimized linear combination of these four RG-invariants.
    The results for the covariance-optimized FSS of $\chi$, $d(\xi / L)/d\beta$, $dU_4/d\beta$, and $dU_6/d\beta$ are taken from Ref.~\cite{PT-11}.
  }
  \begin{ruledtabular}
    \begin{tabular}{l@{}tftto}
      \multicolumn{6}{c}{$\chi$} \\
      \hline
L     & \multicolumn{1}{c}{Fixed $\xi/L$} & \multicolumn{1}{c}{Fixed $Z_a/Z_p$} & \multicolumn{1}{c}{Fixed $U_4$} & \multicolumn{1}{c}{Fixed $U_6$} & \multicolumn{1}{r}{Cov. opt.} \\
\hline
   8  &     5.7     &     2.4     &     2.8     &     3.8     &    11.5     \\
  16  &     3.5     &     3.3     &     1.9     &     2.6     &    18.4     \\
  32  &     3.2     &     4.7     &     1.7     &     2.3     &    27.2     \\
  64  &     2.4     &     2.7     &     1.5     &     2.0     &    13.3     \\
 128  &     2.7     &     4.8     &     1.5     &     2.0     &    22.1     \\
     \multicolumn{6}{c}{$d(\xi / L)/d\beta$} \\
     \hline
L     & \multicolumn{1}{c}{Fixed $\xi/L$} & \multicolumn{1}{c}{Fixed $Z_a/Z_p$} & \multicolumn{1}{c}{Fixed $U_4$} & \multicolumn{1}{c}{Fixed $U_6$} & \multicolumn{1}{r}{Cov. opt.} \\
\hline
   8  &     1.1     &     1.2     &     0.7     &     0.8     &     6.0     \\
  16  &     0.8     &     1.1     &     0.5     &     0.6     &     4.7     \\
  32  &     0.7     &     1.1     &     0.5     &     0.5     &     4.5     \\
  64  &     0.8     &     1.0     &     0.6     &     0.6     &     2.6     \\
 128  &     0.6     &     1.0     &     0.4     &     0.5     &     2.9     \\
     \multicolumn{6}{c}{$dU_4/d\beta$} \\
     \hline
L     & \multicolumn{1}{c}{Fixed $\xi/L$} & \multicolumn{1}{c}{Fixed $Z_a/Z_p$} & \multicolumn{1}{c}{Fixed $U_4$} & \multicolumn{1}{c}{Fixed $U_6$} & \multicolumn{1}{r}{Cov. opt.} \\
\hline
   8  &     3.0     &     1.6     &     6.7     &     5.1     &    35.4     \\
  16  &     3.0     &     1.5     &     6.2     &     4.8     &    43.6     \\
  32  &     3.5     &     1.7     &     7.0     &     5.4     &    54.9     \\
  64  &     3.2     &     1.5     &     5.8     &     4.7     &    22.9     \\
 128  &     3.7     &     1.6     &     6.8     &     5.4     &    39.8     \\
     \multicolumn{6}{c}{$dU_6/d\beta$} \\
     \hline
L     & \multicolumn{1}{c}{Fixed $\xi/L$} & \multicolumn{1}{c}{Fixed $Z_a/Z_p$} & \multicolumn{1}{c}{Fixed $U_4$} & \multicolumn{1}{c}{Fixed $U_6$} & \multicolumn{1}{r}{Cov. opt.} \\
\hline
   8  &     3.7     &     1.7     &    11.4     &     7.9     &    47.6     \\
  16  &     3.8     &     1.6     &    10.3     &     7.2     &    54.2     \\
  32  &     4.5     &     1.8     &    11.8     &     8.4     &    68.2     \\
  64  &     4.1     &     1.6     &     9.3     &     7.1     &    27.5     \\
 128  &     5.0     &     1.8     &    11.5     &     8.4     &    48.9     \\
     \multicolumn{6}{c}{$d(Z_a / Z_p)/d\beta$} \\
     \hline
L     & \multicolumn{1}{c}{Fixed $\xi/L$} & \multicolumn{1}{c}{Fixed $Z_a/Z_p$} & \multicolumn{1}{c}{Fixed $U_4$} & \multicolumn{1}{c}{Fixed $U_6$} & \multicolumn{1}{r}{Cov. opt.} \\
\hline
   8  &     0.7     &     1.5     &     0.7     &     0.7     &     1.2     \\
  16  &     0.8     &     1.4     &     0.8     &     0.9     &     1.4     \\
  32  &     1.2     &     1.6     &     1.4     &     1.4     &     2.1     \\
  64  &     1.1     &     1.4     &     1.2     &     1.3     &     1.7     \\
 128  &     1.5     &     1.5     &     1.8     &     1.8     &     2.2     \\
    \end{tabular}
  \end{ruledtabular}
  \label{results_2d_metropolis}
\end{table}

\begin{table}\renewcommand*{\arraystretch}{1.4}
  \caption{Same as Table \ref{results_2d_metropolis} for a Wolff dynamics.}
  \begin{ruledtabular}
    \begin{tabular}{l@{}tftto}
     \multicolumn{6}{c}{$\chi$} \\
     \hline
L     & \multicolumn{1}{c}{Fixed $\xi/L$} & \multicolumn{1}{c}{Fixed $Z_a/Z_p$} & \multicolumn{1}{c}{Fixed $U_4$} & \multicolumn{1}{c}{Fixed $U_6$} & \multicolumn{1}{r}{Cov. opt.} \\
\hline
   8  &     3.1     &     0.7     &     1.8     &     2.3     &     5.9     \\
  16  &     3.1     &     0.8     &     1.7     &     2.1     &     5.4     \\
  32  &     2.9     &     0.9     &     1.5     &     1.8     &     5.9     \\
  64  &     3.4     &     1.2     &     1.9     &     2.3     &     6.5     \\
 128  &     4.3     &     1.3     &     1.8     &     2.2     &     7.7     \\
     \multicolumn{6}{c}{$d(\xi / L)/d\beta$} \\
     \hline
L     & \multicolumn{1}{c}{Fixed $\xi/L$} & \multicolumn{1}{c}{Fixed $Z_a/Z_p$} & \multicolumn{1}{c}{Fixed $U_4$} & \multicolumn{1}{c}{Fixed $U_6$} & \multicolumn{1}{r}{Cov. opt.} \\
\hline
   8  &     1.4     &     0.8     &     0.8     &     0.8     &     2.5     \\
  16  &     1.2     &     0.9     &     0.7     &     0.8     &     2.2     \\
  32  &     1.0     &     0.9     &     0.7     &     0.7     &     2.4     \\
  64  &     1.1     &     0.9     &     0.7     &     0.8     &     2.1     \\
 128  &     1.0     &     0.9     &     0.7     &     0.7     &     2.0     \\
     \multicolumn{6}{c}{$dU_4/d\beta$} \\
     \hline
L     & \multicolumn{1}{c}{Fixed $\xi/L$} & \multicolumn{1}{c}{Fixed $Z_a/Z_p$} & \multicolumn{1}{c}{Fixed $U_4$} & \multicolumn{1}{c}{Fixed $U_6$} & \multicolumn{1}{r}{Cov. opt.} \\
\hline
   8  &     1.5     &     1.0     &     3.4     &     2.7     &     6.3     \\
  16  &     1.5     &     1.1     &     3.1     &     2.6     &     8.4     \\
  32  &     1.5     &     1.1     &     2.7     &     2.4     &     7.6     \\
  64  &     1.5     &     1.2     &     2.6     &     2.4     &     5.7     \\
 128  &     1.5     &     1.3     &     2.4     &     2.2     &     5.6     \\
     \multicolumn{6}{c}{$dU_6/d\beta$} \\
     \hline
L     & \multicolumn{1}{c}{Fixed $\xi/L$} & \multicolumn{1}{c}{Fixed $Z_a/Z_p$} & \multicolumn{1}{c}{Fixed $U_4$} & \multicolumn{1}{c}{Fixed $U_6$} & \multicolumn{1}{r}{Cov. opt.} \\
\hline
   8  &     1.6     &     1.0     &     5.1     &     3.9     &     8.5     \\
  16  &     1.7     &     1.2     &     4.4     &     3.5     &     9.6     \\
  32  &     1.6     &     1.1     &     3.6     &     3.1     &     8.7     \\
  64  &     1.6     &     1.3     &     3.4     &     3.0     &     6.7     \\
 128  &     1.7     &     1.3     &     3.1     &     2.8     &     6.5     \\
     \multicolumn{6}{c}{$d(Z_a / Z_p)/d\beta$} \\
     \hline
L     & \multicolumn{1}{c}{Fixed $\xi/L$} & \multicolumn{1}{c}{Fixed $Z_a/Z_p$} & \multicolumn{1}{c}{Fixed $U_4$} & \multicolumn{1}{c}{Fixed $U_6$} & \multicolumn{1}{r}{Cov. opt.} \\
\hline
   8  &     0.8     &     1.5     &     0.8     &     0.8     &     1.2     \\
  16  &     0.9     &     1.3     &     1.0     &     1.0     &     1.4     \\
  32  &     1.0     &     1.1     &     1.1     &     1.1     &     1.3     \\
  64  &     1.0     &     1.2     &     1.2     &     1.2     &     1.4     \\
 128  &     1.1     &     1.2     &     1.3     &     1.3     &     1.6     \\
    \end{tabular}
  \end{ruledtabular}
  \label{results_2d_wolff}
\end{table}

In Table \ref{results_2d_metropolis} we report our benchmarks obtained by sampling the model with the standard Metropolis algorithm.
The CPU gains appear to depend strongly on the observable, and for each observable, there is considerable dependence on the chosen RG-invariant $R$ to be fixed.
In particular, for $\chi$, $dU_4/d\beta$ and $dU_6/d\beta$, we observe that at least one of the four possible choices $R$ gives a significant reduction of the error bars and, consequently, an effective gain in CPU time.
Moreover,
the use of the covariance-optimized RG-invariant greatly improve further the CPU gains, in particular for $dU_4/d\beta$ and $dU_6/d\beta$ for which we observe a rather large CPU gain.
For the observables $d(\xi / L)/d\beta$, $d(Z_a / Z_p)/d\beta$ the CPU gain obtained choosing one of the four RG-invariants is instead limited, and in some cases it is less than 1, meaning that there is actually a (small) increase of the error bars, compared to the standard analysis.
Nevertheless, also for these observables the choice of the covariance-optimized RG-invariant provides a non-negligible gain in CPU time.
Remarkably, the CPU gains are stable on increasing the lattice size.
This especially underscores the effectiveness of the FSS method at fixed RG-invariant, for a reduction of error bars at small lattice sizes only would severely limit the usefulness of (any) improvement method.
In Table \ref{results_2d_wolff} we report our results for the Wolff dynamics.
On a qualitative level, the observed pattern in the CPU gains is very similar to the case of a Metropolis dynamics.
Overall, the reduction of error bars is however smaller than in the former case.

\begin{table}\renewcommand*{\arraystretch}{1.4}
  \caption{Same as Table \ref{results_2d_metropolis} for $d=3$.}
  \begin{ruledtabular}
    \begin{tabular}{l@{}tftto}
     \multicolumn{6}{c}{$\chi$} \\
     \hline
L     & \multicolumn{1}{c}{Fixed $\xi/L$} & \multicolumn{1}{c}{Fixed $Z_a/Z_p$} & \multicolumn{1}{c}{Fixed $U_4$} & \multicolumn{1}{c}{Fixed $U_6$} & \multicolumn{1}{r}{Cov. opt.} \\
\hline
   8  &     9.2     &     4.5     &     1.8     &     2.2     &    21.2     \\
  16  &     9.4     &     6.9     &     1.7     &     2.1     &    27.9     \\
  32  &     8.2     &     7.7     &     1.3     &     1.6     &    26.9     \\
  64  &     8.6     &     6.3     &     1.4     &     1.7     &    22.2     \\
 128  &     9.1     &     8.3     &     1.4     &     1.7     &    25.5     \\
     \multicolumn{6}{c}{$d(\xi / L)/d\beta$} \\
     \hline
L     & \multicolumn{1}{c}{Fixed $\xi/L$} & \multicolumn{1}{c}{Fixed $Z_a/Z_p$} & \multicolumn{1}{c}{Fixed $U_4$} & \multicolumn{1}{c}{Fixed $U_6$} & \multicolumn{1}{r}{Cov. opt.} \\
\hline
   8  &     1.2     &     1.1     &     0.6     &     0.6     &     4.2     \\
  16  &     1.1     &     1.1     &     0.5     &     0.6     &     4.9     \\
  32  &     1.0     &     1.0     &     0.4     &     0.5     &     6.2     \\
  64  &     0.8     &     0.9     &     0.4     &     0.4     &     4.2     \\
 128  &     1.0     &     1.1     &     0.4     &     0.5     &     6.1     \\
     \multicolumn{6}{c}{$dU_4/d\beta$} \\
     \hline
L     & \multicolumn{1}{c}{Fixed $\xi/L$} & \multicolumn{1}{c}{Fixed $Z_a/Z_p$} & \multicolumn{1}{c}{Fixed $U_4$} & \multicolumn{1}{c}{Fixed $U_6$} & \multicolumn{1}{r}{Cov. opt.} \\
\hline
   8  &     1.3     &     1.2     &     1.5     &     1.4     &    24.2     \\
  16  &     1.3     &     1.2     &     1.5     &     1.4     &     9.1     \\
  32  &     1.2     &     1.2     &     1.5     &     1.4     &    23.0     \\
  64  &     1.3     &     1.2     &     1.7     &     1.6     &    16.1     \\
 128  &     1.2     &     1.2     &     1.5     &     1.5     &    19.7     \\
     \multicolumn{6}{c}{$dU_6/d\beta$} \\
     \hline
L     & \multicolumn{1}{c}{Fixed $\xi/L$} & \multicolumn{1}{c}{Fixed $Z_a/Z_p$} & \multicolumn{1}{c}{Fixed $U_4$} & \multicolumn{1}{c}{Fixed $U_6$} & \multicolumn{1}{r}{Cov. opt.} \\
\hline
   8  &     1.8     &     1.5     &     3.1     &     2.5     &    27.6     \\
  16  &     1.7     &     1.5     &     3.0     &     2.5     &    24.2     \\
  32  &     1.7     &     1.5     &     3.0     &     2.6     &    25.8     \\
  64  &     1.9     &     1.6     &     3.5     &     2.9     &    25.2     \\
 128  &     1.6     &     1.4     &     3.1     &     2.7     &    23.0     \\
     \multicolumn{6}{c}{$d(Z_a / Z_p)/d\beta$} \\
     \hline
L     & \multicolumn{1}{c}{Fixed $\xi/L$} & \multicolumn{1}{c}{Fixed $Z_a/Z_p$} & \multicolumn{1}{c}{Fixed $U_4$} & \multicolumn{1}{c}{Fixed $U_6$} & \multicolumn{1}{r}{Cov. opt.} \\
\hline
   8  &     1.8     &     1.3     &     1.0     &     1.1     &     2.2     \\
  16  &     1.4     &     1.2     &     0.7     &     0.7     &     2.2     \\
  32  &     1.1     &     1.1     &     0.5     &     0.6     &     2.4     \\
  64  &     1.0     &     1.0     &     0.5     &     0.6     &     1.9     \\
 128  &     0.9     &     1.0     &     0.5     &     0.5     &     2.0     \\
    \end{tabular}
  \end{ruledtabular}
  \label{results_3d_metropolis}
\end{table}

\begin{table}\renewcommand*{\arraystretch}{1.4}
  \caption{Same as Table \ref{results_2d_wolff} for $d=3$.}
  \begin{ruledtabular}
    \begin{tabular}{l@{}tftto}
     \multicolumn{6}{c}{$\chi$} \\
     \hline
L     & \multicolumn{1}{c}{Fixed $\xi/L$} & \multicolumn{1}{c}{Fixed $Z_a/Z_p$} & \multicolumn{1}{c}{Fixed $U_4$} & \multicolumn{1}{c}{Fixed $U_6$} & \multicolumn{1}{r}{Cov. opt.} \\
\hline
   8  &     4.2     &     1.0     &     1.0     &     1.1     &     6.3     \\
  16  &     4.4     &     1.2     &     1.0     &     1.1     &     7.0     \\
  32  &     5.0     &     1.4     &     1.2     &     1.3     &     7.8     \\
  64  &     5.7     &     1.7     &     1.1     &     1.1     &     9.5     \\
 128  &     6.5     &     2.1     &     1.2     &     1.3     &    10.1     \\
     \multicolumn{6}{c}{$d(\xi / L)/d\beta$} \\
     \hline
L     & \multicolumn{1}{c}{Fixed $\xi/L$} & \multicolumn{1}{c}{Fixed $Z_a/Z_p$} & \multicolumn{1}{c}{Fixed $U_4$} & \multicolumn{1}{c}{Fixed $U_6$} & \multicolumn{1}{r}{Cov. opt.} \\
\hline
   8  &     1.2     &     0.9     &     0.6     &     0.6     &     2.7     \\
  16  &     1.1     &     0.8     &     0.5     &     0.5     &     3.1     \\
  32  &     1.2     &     0.8     &     0.6     &     0.6     &     3.1     \\
  64  &     1.1     &     0.8     &     0.5     &     0.5     &     3.2     \\
 128  &     1.1     &     0.9     &     0.5     &     0.6     &     3.1     \\
     \multicolumn{6}{c}{$dU_4/d\beta$} \\
     \hline
L     & \multicolumn{1}{c}{Fixed $\xi/L$} & \multicolumn{1}{c}{Fixed $Z_a/Z_p$} & \multicolumn{1}{c}{Fixed $U_4$} & \multicolumn{1}{c}{Fixed $U_6$} & \multicolumn{1}{r}{Cov. opt.} \\
\hline
   8  &     1.1     &     1.1     &     1.2     &     1.1     &     7.0     \\
  16  &     1.1     &     1.1     &     1.2     &     1.2     &     5.4     \\
  32  &     1.1     &     1.1     &     1.2     &     1.1     &     4.2     \\
  64  &     1.1     &     1.1     &     1.3     &     1.2     &     4.2     \\
 128  &     1.0     &     1.1     &     1.2     &     1.2     &     4.6     \\
     \multicolumn{6}{c}{$dU_6/d\beta$} \\
     \hline
L     & \multicolumn{1}{c}{Fixed $\xi/L$} & \multicolumn{1}{c}{Fixed $Z_a/Z_p$} & \multicolumn{1}{c}{Fixed $U_4$} & \multicolumn{1}{c}{Fixed $U_6$} & \multicolumn{1}{r}{Cov. opt.} \\
\hline
   8  &     1.3     &     1.2     &     1.9     &     1.6     &     7.7     \\
  16  &     1.3     &     1.2     &     2.0     &     1.8     &     6.7     \\
  32  &     1.2     &     1.2     &     1.8     &     1.6     &     5.5     \\
  64  &     1.2     &     1.2     &     1.9     &     1.7     &     6.3     \\
 128  &     1.2     &     1.2     &     1.9     &     1.7     &     5.8     \\
     \multicolumn{6}{c}{$d(Z_a / Z_p)/d\beta$} \\
     \hline
L     & \multicolumn{1}{c}{Fixed $\xi/L$} & \multicolumn{1}{c}{Fixed $Z_a/Z_p$} & \multicolumn{1}{c}{Fixed $U_4$} & \multicolumn{1}{c}{Fixed $U_6$} & \multicolumn{1}{r}{Cov. opt.} \\
\hline
   8  &     1.2     &     1.0     &     0.8     &     0.8     &     1.4     \\
  16  &     1.1     &     1.0     &     0.8     &     0.8     &     1.4     \\
  32  &     1.1     &     0.9     &     0.7     &     0.8     &     1.4     \\
  64  &     1.0     &     0.9     &     0.7     &     0.7     &     1.4     \\
 128  &     1.0     &     0.9     &     0.7     &     0.7     &     1.4     \\
    \end{tabular}
  \end{ruledtabular}
  \label{results_3d_wolff}
\end{table}

\begin{table*}\renewcommand*{\arraystretch}{1.4}
  \caption{Benchmark of the statistical precision of $\beta_f(L)$ as obtained by a FSS analysis at fixed RG-invariant $R$, with $R=\xi/L$, $U_4$, $U_6$, $Z_a/Z_p$, as well as choosing $R$ as the covariance-optimized linear combination of these four RG-invariants.}
  \begin{ruledtabular}
    \begin{tabular}{l@{}ccccc}
     \multicolumn{6}{c}{Ising 2D Metropolis dynamics} \\
     \hline
L     & \multicolumn{1}{c}{Fixed $\xi/L$} & \multicolumn{1}{c}{Fixed $Z_a/Z_p$} & \multicolumn{1}{c}{Fixed $U_4$} & \multicolumn{1}{c}{Fixed $U_6$} & \multicolumn{1}{r}{Cov. opt.} \\
\hline
   8  &    \np{4.2e-4}     &    \np{2.5e-4}     &    \np{4.6e-4}     &    \np{4.4e-4}     &    \np{2.5e-4}     \\
  16  &    \np{4.0e-4}     &    \np{2.1e-4}     &    \np{4.4e-4}     &    \np{4.1e-4}     &    \np{1.8e-4}     \\
  32  &    \np{2.4e-4}     &    \np{1.3e-4}     &    \np{2.7e-4}     &    \np{2.6e-4}     &    \np{1.0e-4}     \\
  64  &    \np{8.6e-5}     &    \np{4.5e-5}     &    \np{9.4e-5}     &    \np{8.8e-5}     &    \np{4.3e-5}     \\
 128  &    \np{5.8e-5}     &    \np{3.0e-5}     &    \np{6.4e-5}     &    \np{6.1e-5}     &    \np{2.3e-5}     \\
     \multicolumn{6}{c}{Ising 2D Wolff dynamics} \\
     \hline
L     & \multicolumn{1}{c}{Fixed $\xi/L$} & \multicolumn{1}{c}{Fixed $Z_a/Z_p$} & \multicolumn{1}{c}{Fixed $U_4$} & \multicolumn{1}{c}{Fixed $U_6$} & \multicolumn{1}{r}{Cov. opt.} \\
\hline
   8  &    \np{1.9e-4}     &    \np{1.8e-4}     &    \np{2.0e-4}     &    \np{1.9e-4}     &    \np{1.5e-4}     \\
  16  &    \np{1.3e-4}     &    \np{1.3e-4}     &    \np{1.5e-4}     &    \np{1.4e-4}     &    \np{1.1e-4}     \\
  32  &    \np{7.1e-5}     &    \np{6.5e-5}     &    \np{8.1e-5}     &    \np{7.8e-5}     &    \np{5.7e-5}     \\
  64  &    \np{3.7e-5}     &    \np{3.7e-5}     &    \np{4.3e-5}     &    \np{4.1e-5}     &    \np{3.2e-5}     \\
 128  &    \np{2.0e-5}     &    \np{1.9e-5}     &    \np{2.4e-5}     &    \np{2.3e-5}     &    \np{1.7e-5}     \\
     \multicolumn{6}{c}{Ising 3D Metropolis dynamics} \\
     \hline
L     & \multicolumn{1}{c}{Fixed $\xi/L$} & \multicolumn{1}{c}{Fixed $Z_a/Z_p$} & \multicolumn{1}{c}{Fixed $U_4$} & \multicolumn{1}{c}{Fixed $U_6$} & \multicolumn{1}{r}{Cov. opt.} \\
\hline
   8  &    \np{1.2e-4}     &    \np{8.1e-5}     &    \np{1.5e-4}     &    \np{1.4e-4}     &    \np{7.8e-5}     \\
  16  &    \np{4.6e-5}     &    \np{3.2e-5}     &    \np{5.8e-5}     &    \np{5.6e-5}     &    \np{2.7e-5}     \\
  32  &    \np{1.6e-5}     &    \np{1.2e-5}     &    \np{2.1e-5}     &    \np{2.0e-5}     &    \np{1.0e-5}     \\
  64  &    \np{4.5e-6}     &    \np{3.4e-6}     &    \np{5.9e-6}     &    \np{5.7e-6}     &    \np{3.0e-6}     \\
 128  &    \np{1.7e-6}     &    \np{1.3e-6}     &    \np{2.2e-6}     &    \np{2.1e-6}     &    \np{1.1e-6}     \\
     \multicolumn{6}{c}{Ising 3D Wolff dynamics} \\
     \hline
L     & \multicolumn{1}{c}{Fixed $\xi/L$} & \multicolumn{1}{c}{Fixed $Z_a/Z_p$} & \multicolumn{1}{c}{Fixed $U_4$} & \multicolumn{1}{c}{Fixed $U_6$} & \multicolumn{1}{r}{Cov. opt.} \\
\hline
   8  &    \np{4.6e-5}     &    \np{4.6e-5}     &    \np{6.5e-5}     &    \np{6.3e-5}     &    \np{4.1e-5}     \\
  16  &    \np{1.6e-5}     &    \np{1.7e-5}     &    \np{2.2e-5}     &    \np{2.2e-5}     &    \np{1.5e-5}     \\
  32  &    \np{5.6e-6}     &    \np{5.8e-6}     &    \np{7.3e-6}     &    \np{7.1e-6}     &    \np{5.2e-6}     \\
  64  &    \np{2.0e-6}     &    \np{2.1e-6}     &    \np{2.6e-6}     &    \np{2.6e-6}     &    \np{1.9e-6}     \\
 128  &    \np{7.0e-7}     &    \np{7.2e-7}     &    \np{9.3e-7}     &    \np{9.2e-7}     &    \np{6.6e-7}     \\
    \end{tabular}
  \end{ruledtabular}
  \label{results_betaf}
\end{table*}
Simulations in $d=3$ have been carried out at $\beta_\text{run}=0.2216544$, close to the accurate estimate of Ref.~\cite{FXL-18} of the critical inverse temperature $\beta_c= \numprint{0.221654626}(5)$.
For the FSS analysis at fixed RG-invariant, we have employed the critical-point values
$(\xi/L)^*=0.6431(1)$, $U_4^*=1.6036(1)$, $U_6^*=3.1053(5)$, $(Z_a / Z_p)^*=0.5425(1)$ \cite{Hasenbusch-10}.
The results of the benchmarks are reported in Table \ref{results_3d_metropolis} for a Metropolis sampling and in Table \ref{results_3d_wolff} for a Wolff dynamics
\footnote{Due to a trivial mistake, the CPU gains reported in Table II of Ref.~\cite{PT-11} for the covariance-optimized FSS
of the three-dimensional Ising model at $L=8$, sampled with Metropolis dynamics, are underestimated by a factor 2.
Also, due to a typo in the analysis, in the case of the three-dimensional Ising model and Wolff dynamics there is a small numerical error in the values of $dU_4/d\beta$ at fixed covariance-optimized RG-invariant,
as reported in Table II of Ref.~\cite{PT-11}.
In both cases, Table \ref{results_3d_wolff} here contains the corrected values.
}.
A comparison with the results for $d=2$ (Tables \ref{results_2d_metropolis} and \ref{results_2d_wolff}) reveals a striking similarity.
For each observable, there is a choice of RG-invariant to be fixed, among $\xi/L$, $Z_a/Z_p$, $U_4$, $U_6$, where the CPU gain is maximized.
Such an optimal choice appears, for a given observable, to be the same across $d=2,3$, and for the two dynamics tested here.
The covariance-optimized linear combination of RG-invariants provides a further important gain in error bars, in line with the results of $d=2$.
The fact that these
observations hold irrespective of dimension and sampling algorithm,
hints at a possible underlying physical mechanism explaining the significant statistical covariance between observables; this, indeed, according to Eq.~(\ref{var_Of}) is ultimately responsible for the reduction of error bars.

Another quantity of interest is the pseudocritical inverse temperature $\beta_f(L)$, which converges to $\beta_c$ for $L\rightarrow\infty$ [see Eqs.~(\ref{betaf_vs_betac_gen}) and (\ref{betaf_vs_betac_crit})].
In Table~\ref{results_betaf} we report the statistical error bars on $\beta_f(L)$ as obtained by a FSS at fixed RG-invariant, for the same MC data used in the previous benchmarks.
We also test the idea of a covariance-optimized linear combination of $\{R_i\}$ that minimizes the error bar in $\beta_f(L)$ [Eq.~(\ref{optimal_lambda_betaf})]; analogous to the benchmarks in Tables~\ref{results_2d_metropolis}-\ref{results_3d_wolff}, the coefficients are fixed by optimizing $\beta_f(L)$ for the largest available lattice size $L=128$.
In analyzing the results in Table~\ref{results_betaf}, it is not particularly meaningful to compare numbers between different lattice size: indeed, the resulting precision in $\beta_f(L)$ depends on the size of the MC sample.
Inspecting Table \ref{results_betaf}, we observe that generically the precision in $\beta_f(L)$ fixing $\xi/L$ or $Z_a / Z_p$ is only slightly better than obtained by choosing $R=U_4$ or $R=U_6$.
The covariance-optimized choice of $R$ appears to give an appreciable reduction of error bar only for the case of the 2D Ising model sampled with a Metropolis dynamics; in the other cases the improvement is limited.

\section{Improved three-dimensional $XY$ model}
\label{sec:xy}
We simulate the O(2) $\phi^4$ model on a three-dimensional lattice of linear size $L$, applying periodic BCs along all directions.
The reduced Hamiltonian ${\cal H}$, such that the Gibbs weight is $\exp(-\cal H)$, is
\begin{equation}
  {\cal H} = -\beta\sum_{\< i\ j\>}\vec{\phi}_i\cdot\vec{\phi}_j
  +\sum_i[\vec{\phi}_i^{\,2}+\lambda(\vec{\phi}_i^{\,2}-1)^2],
  \label{phi4}
\end{equation}
where $\vec{\phi}_x$ is a $2-$component real field on the lattice site $x$, the first sum extends over the nearest-neighbor pairs, and the last term is a sum over all lattice sites.
MC simulations have been carried out combining Metropolis, overrelaxation, and Wolff single-cluster updates \cite{Wolff-89}; technical details of the sampling algorithm can be found in Ref.~\cite{PT-20}.
In the limit $\lambda\rightarrow\infty$ the Hamiltonian (\ref{phi4}) reduces to the standard $XY$ model.
Starting from $\lambda=\infty$, the $(\beta,\lambda)$ phase diagram exhibits a line of continuous phase transitions in the classical $XY$ universality class.
For $\lambda=2.15(5)$ \cite{CHPV-06} the model is {\it improved}, i.e., the leading irrelevant scaling field with dimension $y_i=-0.789(4)$ \cite{Hasenbusch-19} is suppressed.
Improved models are instrumental in high-precision numerical studies of critical phenomena \cite{PV-02}.
For $\lambda=2.1$ the model is critical at $\beta_c=\numprint{0.5091503}(3)$ and for $\lambda=2.2$, it is critical at $\beta_c=\numprint{0.5083355}(3)$ \cite{CHPV-06}.

\begin{figure}
  \centering
  \includegraphics[width=0.9\linewidth]{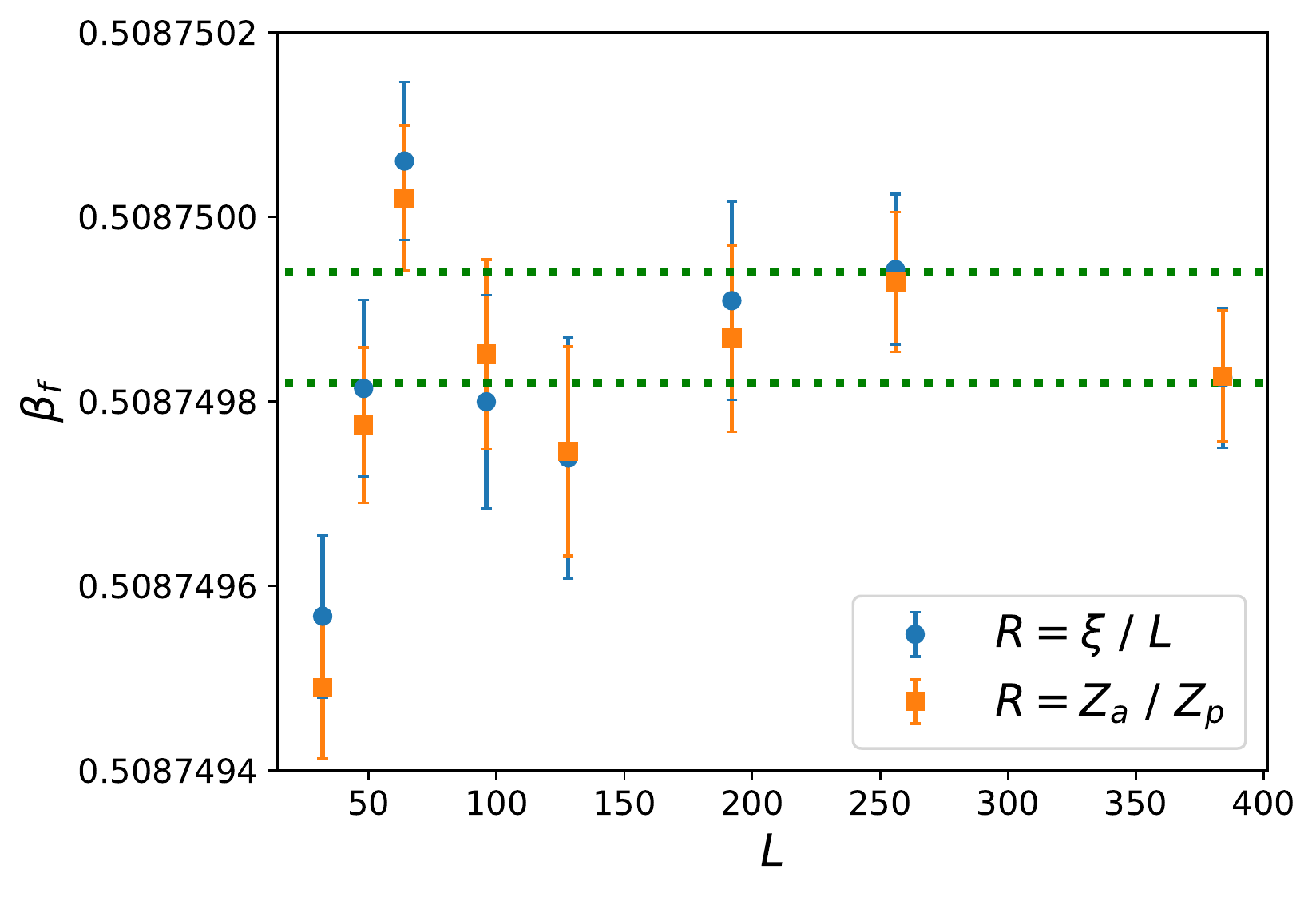}
  \caption{Pseudocritical inverse temperature $\beta_f(L)$ for the improved $XY$ model of Eq.~(\ref{betaf_xy}), as obtained by FSS at fixed $\xi/L$ and fixed $Z_a/Z_p$.
    Dotted lines indicate the final estimate of $\beta_c$ [Eq.~(\ref{betac_xy})].}
  \label{betaf_xy}
\end{figure}
Here, we have used FSS at fixed RG to compute the critical value of $\beta_c$ for the estimated central value of $\lambda=2.15$ at which the model is improved.
To this end, we have simulated the model (\ref{phi4}) for lattice sizes $L=32-384$ with high statistics, and analyzed the data at fixed $\xi/L$ and $Z_a / Z_p$, employing the fixed-point values $(\xi/L)^*=0.59238(7)$ and $(Z_a/Z_p)^*=0.32037(6)$ \cite{Hasenbusch-19}.
According to Eq.~(\ref{betaf_vs_betac_crit}), we expect an improved approach of $\beta_f(L)$ to the inverse critical temperature.
Furthermore, being the model improved, the leading irrelevant scaling field, associated with the cubic anisotropy of the lattice, has a negative scaling dimension $\omega=\omega_{\rm nr}\approx 2$ \cite{Hasenbusch-19}.
Therefore, we expect a rather fast convergence of $\beta_f(L) \rightarrow \beta_c$ on increasing $L$.
This is confirmed in Fig.~\ref{betaf_xy}, where we show $\beta_f(L)$ as obtained by fixing $\xi/L$ or $Z_a/Z_p$.
For $L\ge 96$, data points of $\beta_f(L)$ are mutually compatible within the error bars.
Accordingly, we can safely average the data for the two largest lattice sizes $L=256$, $384$, so as to obtain a statistically improved value of $\beta_c$.
Doing so for data at fixed $\xi/L$ we obtain $\beta_c=\numprint{0.50874988}(6)$, while using the analysis at fixed $Z_a/Z_p$ we get $\beta_c=\numprint{0.50874988}(5)$.
From these two values, we take as a conservative estimate
\begin{equation}
  \beta_c=\numprint{0.50874988}(6).
  \label{betac_xy}
\end{equation}
Including also the data at $L=192$ in the average does not change the final estimate: in fact, as shown in Fig.~\ref{betaf_xy}, the data points for $L=192$ lie well within one error bar of the estimated $\beta_c$.
This further strengthens the robustness of the final result given in Eq.~(\ref{betac_xy}).

\section{Summary}
\label{sec:summary}
FSS at fixed RG-invariant is a convenient and powerful method to analyze MC data of a critical model.
Its implementation, discussed in detail in Sec.~\ref{sec:fss}, essentially consists in a specific analysis of MC data, which does not require additional computational costs.
The two main advantages of this method over more standard FSS techniques are in the observed statistical improvement of critical observables, and in its convenience for computing critical parameters.

Extensive benchmarks carried out in this work show that the method offers in many cases a significant reduction of statistical error bars, as compared to a standard analysis at fixed Hamiltonian parameters.
The improvement of statistical precision is due to the cross-correlations of MC observables.
An optimization of the method put forward in Ref.~\cite{PT-11}, based on the covariance analysis, allows one to obtain rather large gains in the error bars or, equivalently, in the computational time.
In this context,
covariance analysis has been used to
reduce the statistical error of an observable
by adding a so-called control variate, i.e., a quantity whose mean value vanishes \cite{FMM-09}, and to
determine
the optimal weighted average of critical exponents estimators \cite{WJ-09,WJ-10}.
As a tool to improve statistical precision, FSS at fixed RG-invariant largely outperform the aforementioned methods, especially in the case of the covariance-optimized version \cite{PT-11}.
It is worth mentioning that, in principle, one could combine the present method with those of Refs.~\cite{FMM-09,WJ-09,WJ-10}.

FSS at fixed RG-invariant provides also a convenient way to compute the critical parameters of a model.
As shown in Eqs.~(\ref{betaf_vs_betac_gen}), (\ref{betaf_vs_betac_crit}), and (\ref{fss_obs_fixedR}), within this framework pseudocritical couplings and critical observables acquire a simple size dependence, which is the starting point for fitting critical exponents and critical couplings.
As a concrete application,
we have employed the method to accurately determine the critical inverse temperature [Eq.~(\ref{betac_xy})] of an improved model in the three-dimensional $XY$ universality class.

\begin{acknowledgments}
  F.P.T. is funded by the Deutsche Forschungsgemeinschaft (DFG, German Research Foundation), Project No. 414456783.
  The author thanks U. Wolff for useful communications.
  The author gratefully acknowledges the Gauss Centre for Supercomputing e.V. (\href{www.gauss-centre.eu}{www.gauss-centre.eu}) for funding this project by providing computing time on the GCS Supercomputer SuperMUC-NG at Leibniz Supercomputing Centre (\href{www.lrz.de}{www.lrz.de}), where the simulations of the $\phi^4$ model have been carried out.
\end{acknowledgments}

\appendix
\section*{Appendix: Proof of the formulas for the optimal coefficients}
In this appendix we prove Eq.~(\ref{optimal_lambda_O}) which was reported in Ref.~\cite{PT-11}, and Eq.~(\ref{optimal_lambda_betaf}).
The space of real stochastic variable with null expectation value is a real vector space equipped with a scalar product $<\, , \, >$, defined as the covariance
\begin{equation}
  <\delta X, \delta Y> \equiv {\rm COV}(\delta X, \delta Y) = E[\delta X\delta Y].
  \label{scalar_product}
\end{equation}
In particular, the variance of a stochastic variable is the length square computed with the scalar product of Eq.~(\ref{scalar_product}).
Thus, minimizing the variance is equivalent to the problem of minimizing the length of a vector.

Let us first consider the minimization of $\sigma_{\beta_f}$. Inserting Eq.~(\ref{R_lambda}) into Eq.~(\ref{fluctuations_betaf}), and neglecting the second term on the right-hand side (see the discussion in Sec.~\ref{sec:fss:error}), we obtain the fluctuations of $\beta_f$ as a function of the parameters $\{\lambda_i\}$,
\begin{equation}
  \delta\beta_f = -\frac{\sum_i \lambda_i \delta R_i}{\sum_i \lambda_i {\bf R'}_i},
  \label{fluctuations_betaf_lambda}
\end{equation}
with the definition of Eq.~(\ref{def_MNRp}).
Clearly, the right-hand side of Eq.~(\ref{fluctuations_betaf_lambda}) is invariant under a global rescaling $\lambda_i\rightarrow c \lambda_i$: this simply reflects the trivial equivalence of
fixing $R=R_f$ or $cR=cR_f$.
To minimize the variance, it is useful to fix the normalization of $\lambda_i$ as
\begin{equation}
  \sum_i \lambda_i {\bf R'}_i=1,
  \label{normalization_lambda_betaf}
\end{equation}
such that we need to minimize the length of the vector
\begin{equation}
  \delta\beta_f = -\sum_i \lambda_i \delta R_i
  \label{fluctuations_betaf_constrained}
\end{equation}
subject to the constraint of Eq.~(\ref{normalization_lambda_betaf}).
An immediate solution to this problem is obtained by exploiting a geometrical interpretation.
In the vector space spanned by the $N$ fluctuations $\delta R_i$, the coefficients $-\lambda_i$ of Eq.~(\ref{fluctuations_betaf_constrained}) are the coordinates in the (nonorthogonal) base $\{\delta R_i\}$.
Accordingly, the constraint of Eq.~(\ref{normalization_lambda_betaf}) is an affine subspace of codimension 1, which contains the vectors $\delta v\equiv \lambda_i \delta R_i$ satisfying
\begin{equation}
  \sum_{ij} \lambda_i {\bf M}_{ij} \left({\bf M^{-1}R'}\right)_j = <\delta v, \delta w> = 1,
  \label{affine_betaf}
\end{equation}
where the vector $\delta w$
\begin{equation}
  \delta w = \sum_i \left({\bf M^{-1}R'}\right)_i \delta R_i
  \label{w}
\end{equation}
is orthogonal to the affine subspace of the constraint.
\begin{figure}
  \centering
  \includegraphics[width=0.6\linewidth]{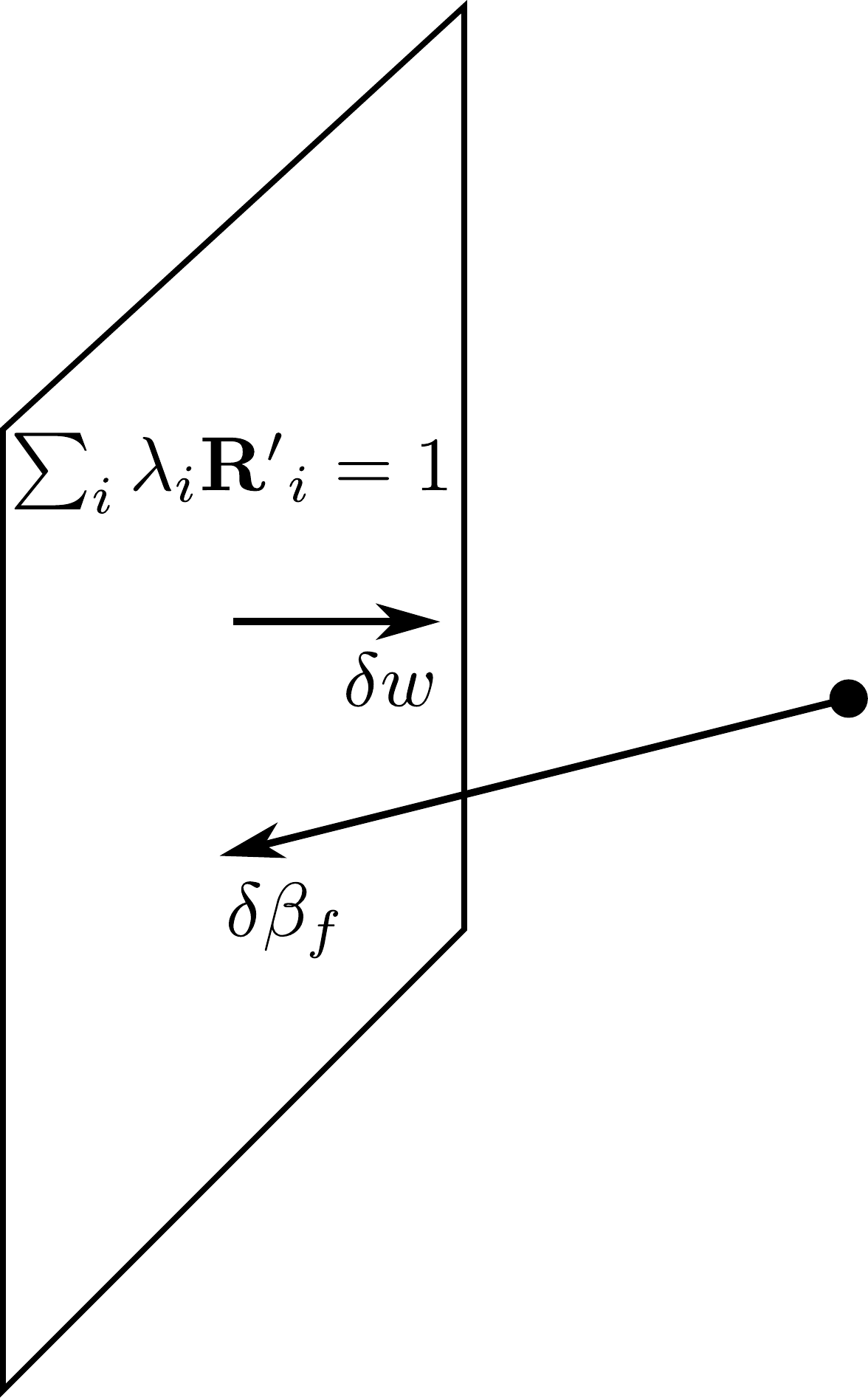}
  \caption{Geometrical illustration of the problem of minimizing the fluctuations of $\beta_f$.
    The hyperplane defined by the constraint of Eq.~(\ref{normalization_lambda_betaf}) is an affine subspace of codimension 1, orthogonal to the vector $\delta w$. The fluctuations $\delta\beta_f$ are a vector from the origin to the hyperplane. Its length is minimized when $\delta\beta_f$ is orthogonal to the hyperplane.}
  \label{geometry_betaf}
\end{figure}
An illustration of the problem is shown in Fig.~\ref{geometry_betaf}.
The minimum length of $\delta \beta_f$ is readily found when $\delta\beta_f$ is orthogonal to the affine subspace, i.e., when
\begin{equation}
  \delta\beta_f = \mu \delta w.
  \label{betaf_mu}
\end{equation}
The coefficient $\mu$ is easily obtained by inserting Eq.~(\ref{betaf_mu}) into Eq.~(\ref{affine_betaf}) or Eq.~(\ref{normalization_lambda_betaf}), obtaining the optimal coefficients $\{\lambda_i\}$ given in Eq.~(\ref{optimal_lambda_betaf}).

To prove Eq.~(\ref{optimal_lambda_O}) one proceeds analogously.
Using Eq.~(\ref{R_lambda}) in Eq.~(\ref{fluctuations_Of}), and keeping only the leading term, we obtain the fluctuations of $O_f$
\begin{equation}
    \delta O_f = \delta O - E\left[\frac{\partial O}{\partial\beta}\right] \frac{\sum_i \lambda_i \delta R_i}{\sum_i \lambda_i {\bf R'}_i}.
  \label{fluctuations_Of_lambda}
\end{equation}
This time we fix the normalization of $\{\lambda_i\}$ as
\begin{equation}
  \sum_i \lambda_i {\bf R'}_i = E\left[\frac{\partial O}{\partial\beta}\right],
  \label{normalization_lambda_Of}
\end{equation}
such that we need to minimize the length of the vector $\delta O_f$,
\begin{equation}
    \delta O_f = \delta O - \sum_i \lambda_i \delta R_i,
  \label{fluctuations_Of_lambda_constrained}
\end{equation}
subject to the constraint of Eq.~(\ref{normalization_lambda_Of}), which represents an affine subspace, parallel to the one of Eq.~(\ref{normalization_lambda_betaf}).
Once again, the minimum length of $\delta O_f$ is found when the right-hand side of Eq.~(\ref{fluctuations_Of_lambda_constrained}) is parallel to $\delta w$ defined in Eq.~(\ref{w}),
\begin{equation}
  \delta O_f = \delta O - \sum_i \lambda_i \delta R_i = \mu \delta{w}.
  \label{Of_mu}
\end{equation}
By taking the scalar product of both sides of Eq.~(\ref{Of_mu}) with $\delta R_i$ we obtain an equation for $\{\lambda_i\}$
\begin{equation}
  {\bf N}_i - \sum_j {\bf M}_{ij}\lambda_j = \mu {\bf R'}_i,
  \label{lambda_mu}
\end{equation}
where we have used Eq.~(\ref{w}) and the definitions of Eq.~(\ref{def_MNRp}).
Equation (\ref{lambda_mu}) can be inverted to obtain
\begin{equation}
  \lambda_i = \left[{\bf M}^{-1}\left({\bf N}-\mu {\bf R'}\right)\right]_i.
  \label{lambda_mu2}
\end{equation}
Inserting Eq.~(\ref{lambda_mu2}) into Eq.~(\ref{normalization_lambda_Of}) we find the value of the constant $\mu$,
\begin{equation}
  \mu = \frac{{\bf R'^T M^{-1}N}-E[\partial O/\partial\beta]}{\bf R'^TM^{-1}R'}.
  \label{mu_solution}
\end{equation}
Finally, using this result in Eq.~(\ref{lambda_mu2}) we obtain Eq.~(\ref{optimal_lambda_O}).

\bibliography{francesco}
\end{document}